\documentclass[12pt, preprint]{aastex}
\usepackage{graphicx}
\usepackage{color}
\usepackage{amssymb,amsmath}
\newcommand{\nside}{N_{\rm side}}
\begin{document}

\title{Isotropy in the two-point angular correlation function of the CMB}

\author{Sophie Zhang}
\affil{Department of Physics, University of Michigan}
\affil{450 Church St, Ann Arbor, MI 48109-1040}
\email{sophz@umich.edu}

\begin{abstract}
We study the directional dependence of the angular two-point correlation
function in maps of the cosmic microwave background (CMB). We propose two new statistics, one which measures the correlation of each point in the sky with a ring
of points separated angle $\theta$ away, and a second that measures
the missing angular correlation above 60 degrees as a function of
direction. Using these statistics, we find that most of the low power in
cut-sky maps measured by the WMAP experiment comes from unusually low
contributions from the directions of the lobes of the quadrupole and the octupole. 
These findings may aid a future explanation of why the CMB exhibits low power at large angular scales.\end{abstract}

\maketitle

\section{Introduction}
	
With the launch of the Wilkinson Microwave Anisotropy Probe (WMAP) in 2003,
the CMB has been measured in highly detailed full-sky maps
\citep{Bennett2003,Spergel:2006,WMAP5,WMAP7}, which have been examined in great detail over
the past few years \citep{Cruz2005,deOliveira2004,Land2005a,Hoftuft:2009,Hansen:2008,KimNaselsky:2010,Bennett:2010}. In particular, the angular two-point correlation function $C(\theta)$
has been studied; it is defined as the average product between the temperature of two points
angle $\theta$ apart
\begin{equation}
C(\theta)=\overline{T(\hat{\Omega}_{1}),T(\hat{\Omega}_{2})} 
|_{\hat{\Omega}_{1} \cdot \hat{\Omega}_{2} = \cos(\theta)} .
\label{eq:Ctheta}
\end{equation}
Here $T(\hat{\Omega})$ is the fluctuation around the mean of the temperature
in direction $\hat\Omega$ on the sky. Several anomalies have been claimed in
the angular correlation function, especially the missing power on large
angular scales (for a review, see \cite{CHSS_review}). Specifically, the
angular correlation function is very nearly zero at scales above $60\degr$;
such a low correlation has a significance of $\geq 3.2 \sigma$ in the standard Gaussian random, statistically isotropic cosmology. This
result, first observed in COBE data, has been strengthened with WMAP
first-year data \citep{DMR4}, as well as later WMAP
observations \citep{wmap123,Copi_nolarge,Sarkar}, though questions have been raised regarding its significance \citep{PeirisPontzen,Efstathiou2010,Ma2011}

In this brief report, we explore the directional contributions to the angular
two-point correlation function. Directional information is
lost in the original definition of $C(\theta)$, as Eq.~(\ref{eq:Ctheta})
assumes statistical isotropy and averages over all directions on the sky. Our
goal is to provide a generalization of Eq.~(\ref{eq:Ctheta}), and ascertain if there are specific directions which result in the unusually low correlation.

\section{Fixed-vertex $C(\theta)$}
	
\begin{figure*}
\includegraphics[scale=0.33]{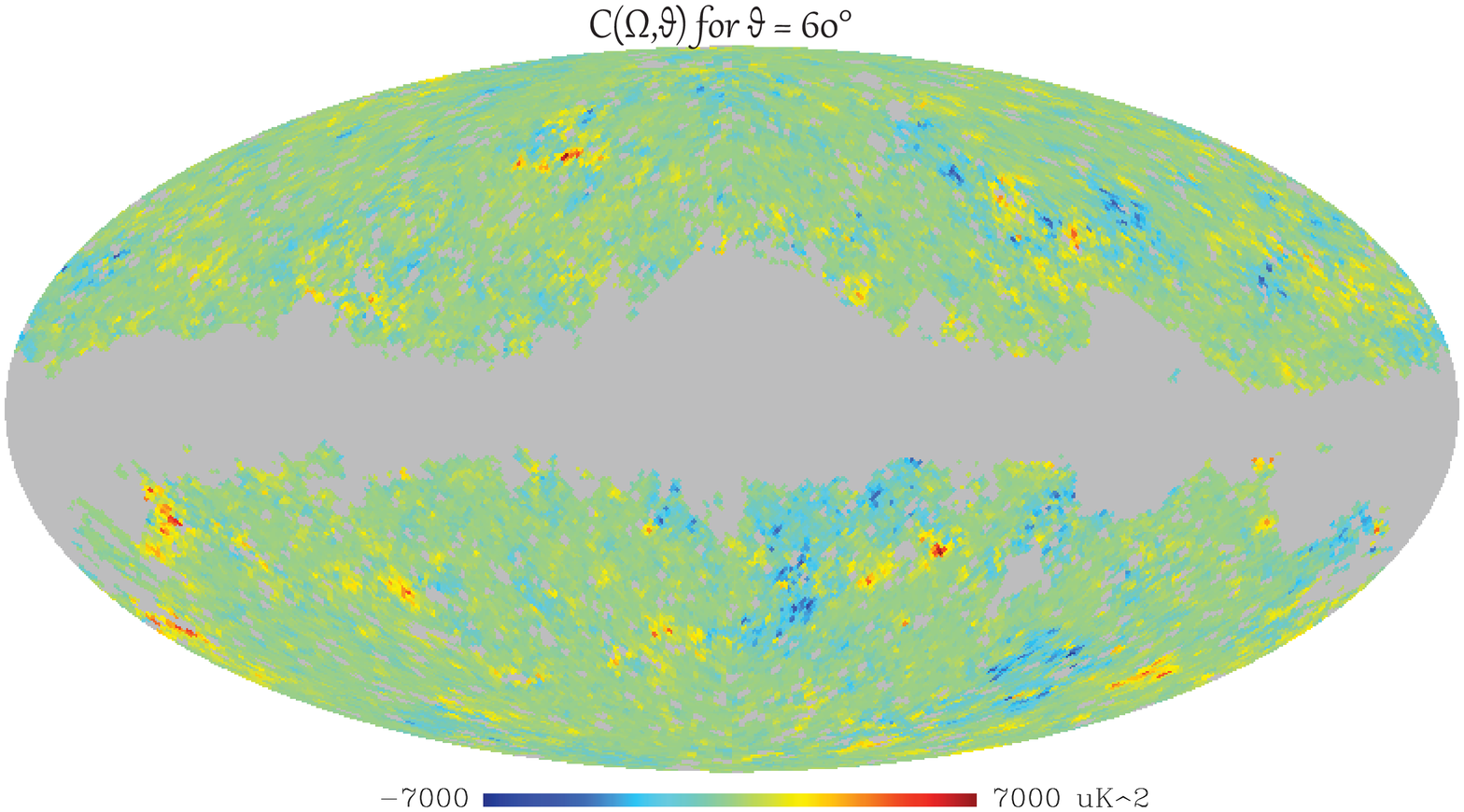}
\includegraphics[scale=0.33]{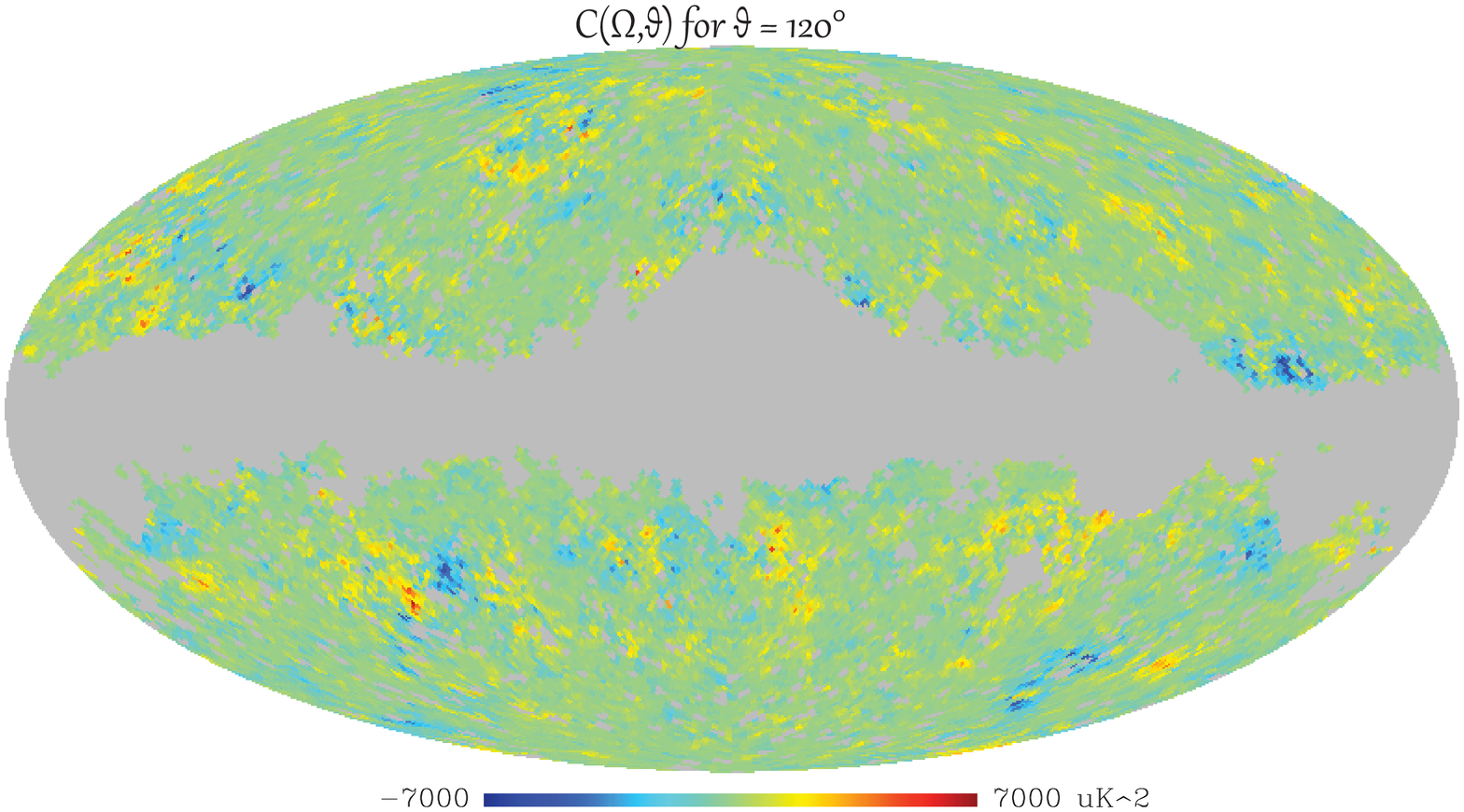}
\caption{Full-sky maps of the fixed-vertex correlation function
  $C(\hat{\Omega},\theta)$ for $\theta = 60\degr$ (left panel) and
  $\theta=120\degr$ (right panel) at resolution $\nside=64$, calculated for the cut-sky ILC map.}
\label{fig:COmega}
\end{figure*}

\begin{figure}
\includegraphics[scale=0.6]{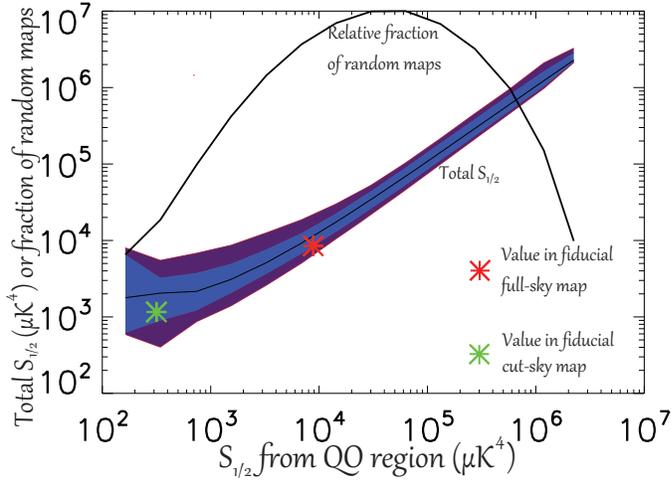}
\caption{The relationship between $S^{\rm QO}_{1/2}$ and $S_{1/2}$ is shown. The first black curve (labeled 'Total $S_{1/2}$') indicates the median relationship in random Gaussian skies. The blue and purple bands surrounding it are (respectively) 68$\%$ and 95$\%$ contours, indicating the spread of the distribution due to cosmic variance. (Note that the contours shown are only approximate for very high or low $S_{1/2}$ values, due to very low statistics.) The second black curve (labeled) logarithmically plots the relative fraction of random maps; the green and red asterisks indicate the values for $S^{\rm QO}_{1/2}$ and $S_{1/2}$ from the fiducial cut-sky and full-sky ILC map, respectively. The narrow contours indicate a very tight relation between the two statistics. The cut sky (green) fits the typical relation, but has an unusually low value for $S_{1/2}^{\rm QO}$; hence, it is possible that $S_{1/2}$ is low because $S_{1/2}^{\rm QO}$ is. See text for details.}
\label{fig:SQO}
\end{figure}

The angular two-point correlation function is defined by Eq.~(\ref{eq:Ctheta}),
where the average is taken over all pairs of points separated by angle
$\theta$ on the sky.

We now propose a new statistical quantity, a fixed-vertex correlation function
\begin{equation}
  C(\hat{\Omega},\theta)=\overline{T(\hat{\Omega}),
    T(\hat{\Omega}_{2})} |_{\hat{\Omega} \cdot \hat{\Omega}_{2} = \cos(\theta)}
\end{equation}
where $\hat\Omega$ is a direction in the sky, and the average is taken over a
ring of pixels separated by $\theta$ from the direction $\hat\Omega$.
In particular, notice that $C(\theta)=\frac{1}{N}\sum_{i=1}^N C(\hat\Omega_i,
\theta)$\footnote{On a masked sky, the standard
  $C(\theta)$ is no longer the simple average of
  $C(\hat{\Omega},\theta)$. We need to introduce the weight function
  $W(\hat{\Omega},\theta)$, the fraction of unmasked points in the ring
  $\theta$ degrees away from $\hat{\Omega}$. Thus,
\begin{equation}
 C(\theta)= \frac{\sum_{i=1}^{N}C(\hat{\Omega_{i}},\theta)W(\hat{\Omega_{i}},\theta)}{\sum_{i=1}^{N}W(\hat{\Omega_{i}},\theta)} \bigg|_{\hat{\Omega}_{1} \cdot \hat{\Omega}_{2} = \cos(\theta)}
\end{equation}
}. Maps of $C(\hat{\Omega},\theta)$ for $\theta$ = $60^\circ$ and $\theta$ = $120^\circ$ are shown in Figure 1.

In order to quantify the low power in the angular two-point function between
approximately $60\degr$ and $180\degr$, the statistic  \citep{Spergel2003}
\begin{equation}
S_{1/2} = \int_{-1}^{1/2}C(\theta)^2 d(\cos \theta)
\end{equation}
has been widely used. In order to examine possible \emph{anisotropies} in
$C(\theta)$, we attempt to separate $S_{1/2}$ into contributions from specific
directions $\hat{\Omega}$; we define
\begin{eqnarray}S_{1/2}(\hat{\Omega}) &\equiv & \frac{1}{N} \int_{-1}^{1/2}C(\hat{\Omega},\theta)C(\theta)d(\cos \theta),
\label{eq:S12Omega_def}
\end{eqnarray}
so that the sum of $S_{1/2}(\hat{\Omega}, \theta)$ over all pixels is
$S_{1/2}$.\footnote{For a masked sky, defining $W(\hat{\Omega}_{i},\theta)$ as the
  fraction of the ring for the respective $C(\hat{\Omega}_{i},\theta)$ that is
  unmasked (W is 0 if $\Omega$ is a masked point)  we have
\begin{eqnarray}
S_{1/2}(\hat{\Omega}_i)&=&\int_{-1}^{1/2}\frac{W(\hat{\Omega}_{i},\theta)C(\hat{\Omega}_{i},\theta)C(\theta)d(\cos
  \theta)}{\sum_{j=1}^{N}W(\hat{\Omega}_{j},\theta)}
\end{eqnarray}
}

Note that the quantity $S_{1/2}(\hat{\Omega})$ is non-local, in that it
receives contributions from not only the direction $\hat\Omega$, but from the
whole sky, due to the term $C(\theta)$ in
Eq.~(\ref{eq:S12Omega_def}). 
So, for example, excluding a pixel $\hat{\Omega}$ will {\it not} change the
value of $S_{1/2}$ simply by $-S_{1/2}(\hat{\Omega})$, because there is also an
indirect effect of the changed global $C(\theta)$ as well as the changed
monopole that is being subtracted. These two are approximately constant if
only a small area of the sky is affected, so analysis of
$S_{1/2}(\hat{\Omega})$ is suitable for examination of the effect of small
regions of the sky on $S_{1/2}$. However, these indirect effects must be taken
into consideration while examining much larger regions.

\begin{figure*}
\includegraphics[scale=0.3]{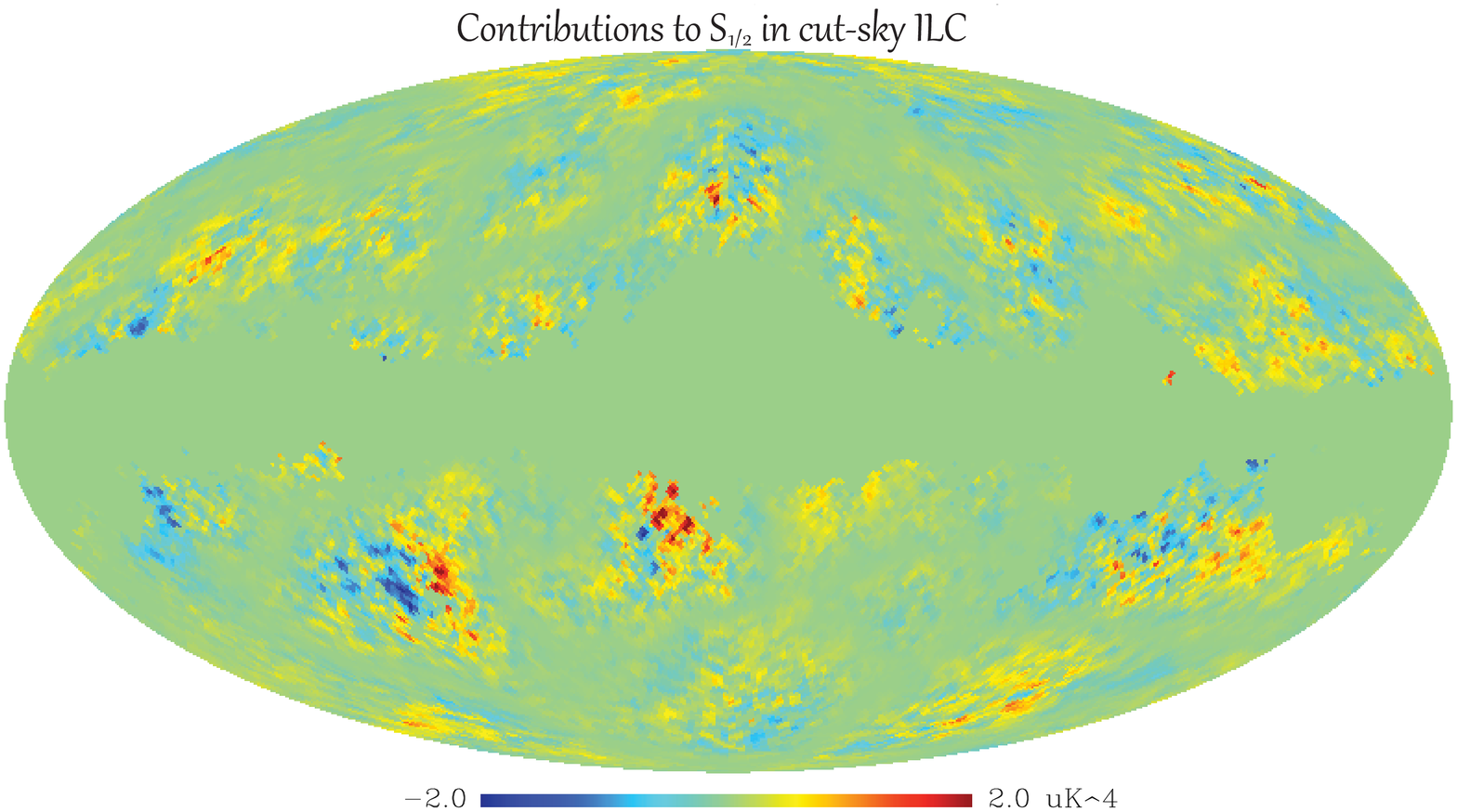}
\includegraphics[scale=0.3]{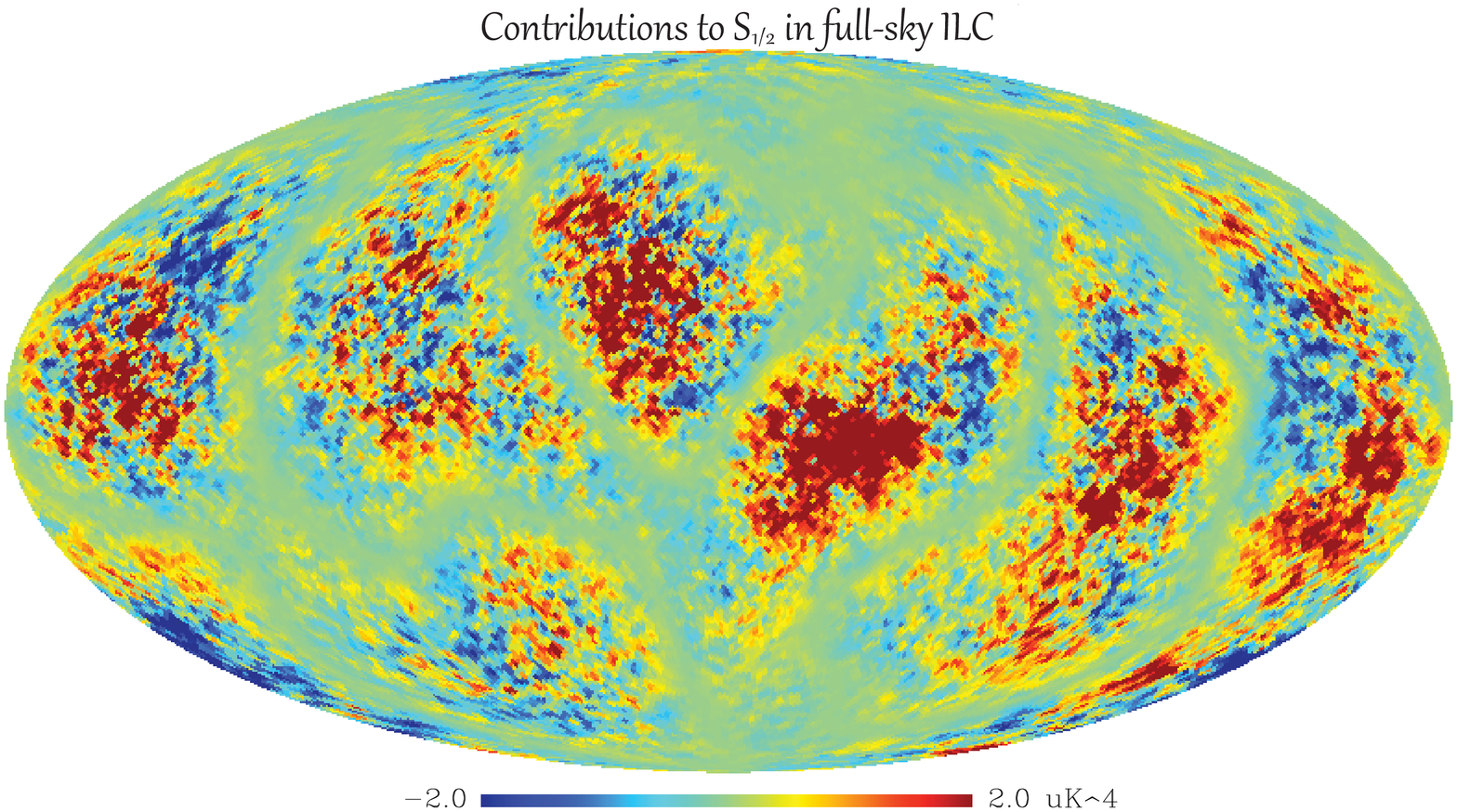}
\caption{The quantity $S_{1/2}(\hat{\Omega})$ defined in Eq (\ref{eq:S12Omega_def}) as a function of direction $\hat{\Omega}$ in the cut-sky ILC map. The left panel shows contributions in the cut-sky map, where the masked pixels contribute nothing and are counted as zero. The right panel shows $S_{1/2}(\hat{\Omega})$ calculated from the full-sky ILC map. In both cases, the total value of $S_{1/2}(\hat{\Omega})$ summed over the entire map is mathematically equal to $S_{1/2}$, which is about 1000 $(\mu \rm{K})^4$ for the cut-sky map, and 8000 $(\mu \rm{K})^4$ for the full-sky map. Note the strong resemblance of the full-sky $S_{1/2}(\hat{\Omega})$ and the power in the quadrupole and octupole (seen in e.g. Fig 4 in \cite{CHSS_review}). }
\label{fig:S12Omega}
\end{figure*}

\begin{figure*}
\includegraphics[scale=0.3]{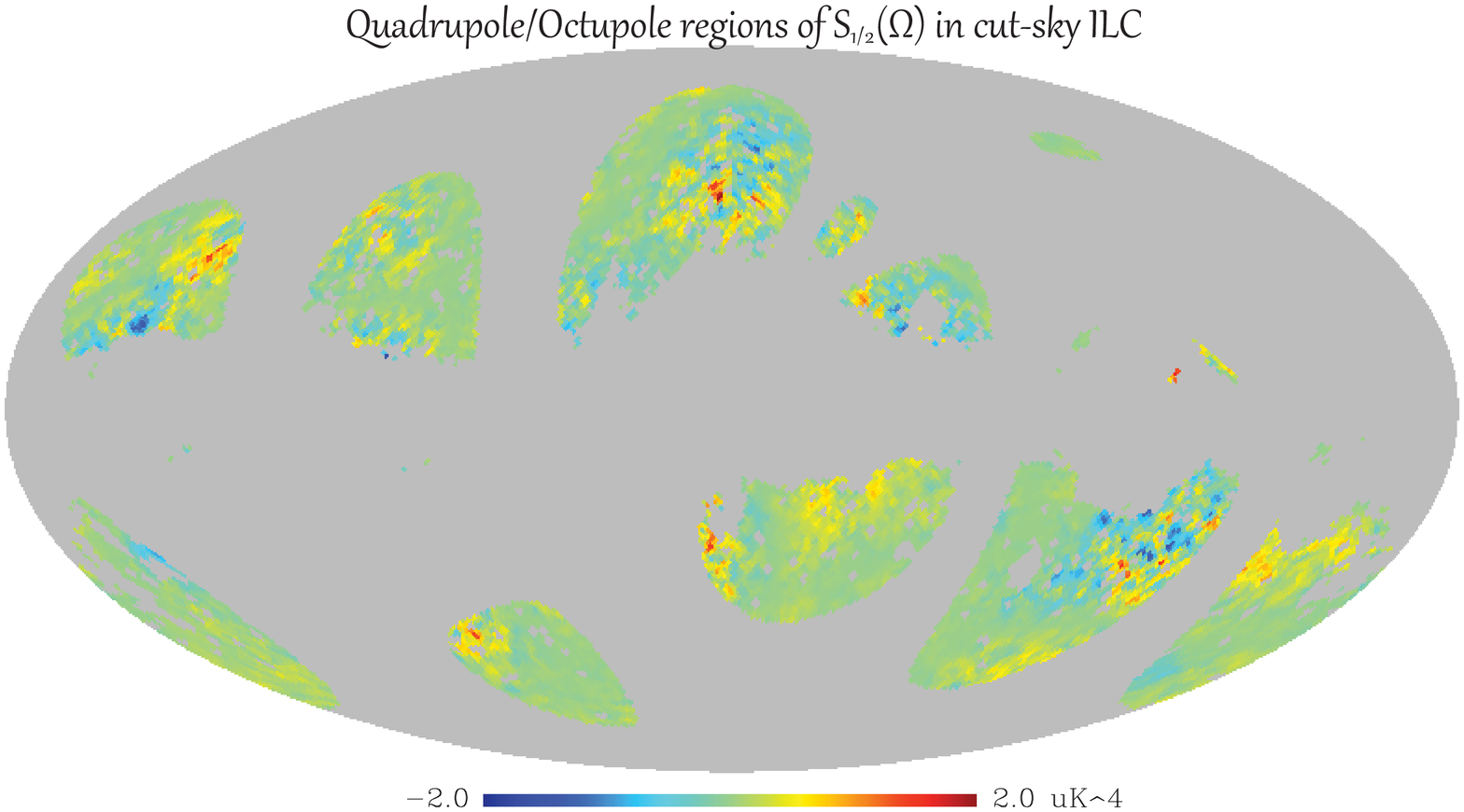}
\includegraphics[scale=0.3]{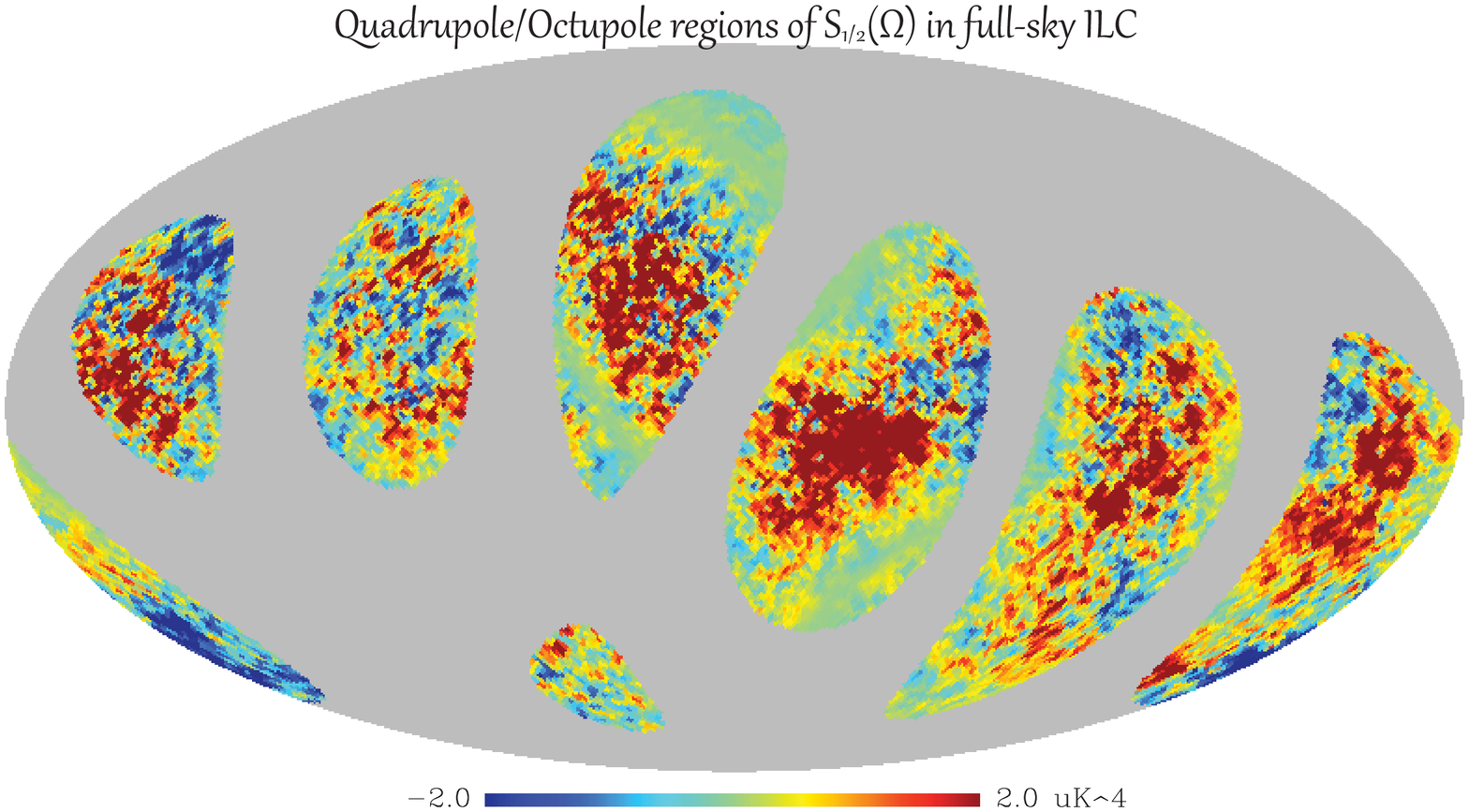}

\caption{$S_{1/2}(\hat{\Omega})$ is plotted in the cut-sky (left panel) and full-sky ILC map (right panel), with only the QO regions (defined in text) shown. Gray indicates both masked points and the non-QO region. In the full-sky map (right), correlation between the structure in $S_{1/2}(\hat{\Omega})$ and the QO regions is apparent, as expected. In the cut-sky map (left), there is little correlation between the structure and the QO regions, a result at odds with typical statistically isotropic random maps. }
  \label{fig:QORegions}
\end{figure*} 
\section{Directional contributions to $S_{1/2}$}
\subsection{Random map analysis}
We first examine plots of $S_{1/2}$ in random maps to obtain an idea of what to expect. We generate 80,000 synthetic Gaussian random statistically isotropic maps, based on the underlying best-fit $\Lambda$CDM cosmological model. \citep{WMAP7Power}

A strong relationship between contributions to power in $S_{1/2}$ and the regions of the quadrupole and octupole (QO regions for simplicity) is consistently seen in random maps. An example may be seen in the right panels of Figures \ref{fig:S12Omega} and \ref{fig:QORegions} with the full-sky ILC map.

In order to quantify these contributions, we repeat our analysis of $S_{1/2}(\hat{\Omega})$ in the regions of the sky
containing only the quadrupole and octupole.  We now compute
$S_{1/2}(\hat{\Omega})$ in a 'sky' containing only the quadrupole and octupole
of each map, and define the QO region to correspond to pixels
$\hat{\Omega}_{QO}$ where $|S_{1/2_{QO}}(\hat{\Omega}_{QO})| \geq x$; $x$ is
selected so that $\sum S_{1/2}(\hat{\Omega}_{QO})/\sum S_{1/2}(\hat{\Omega})
\approx 0.9$. We define the QO region of each individual synthetic map separately. The contribution of $S_{1/2}$ from the QO region is thus defined as 
\begin{equation}
S_{1/2}^{\rm QO}\equiv 
\sum_i S_{1/2}(\hat{\Omega}_i^{\rm QO}).
\end{equation} 

There is an extremely strong, almost linear relationship between $S_{1/2}^{\rm QO}$ and total $S_{1/2}$, plotted in Fig \ref{fig:SQO}. In contrast, the total $S_{1/2}$ and the contributions from non-QO regions of the sky are essentially uncorrelated, except at very high ($\geq 10^5$ $(\mu \rm{K})^4$) values of $S_{1/2}$, with which we are not concerned. Note that by definition, $S_{1/2} \equiv S_{1/2}^{QO} + S_{1/2}^{\rm non-QO}$. 

This result makes intuitive sense, as the statistic $S_{1/2}(\hat{\Omega})$ is based off of $C(\hat{\Omega}_i,\theta)$, which is higher when the central point $\hat{\Omega}$ is in a high-temperature region such as the QO region. More qualitatively, $S_{1/2}(\hat{\Omega})$ is a measure of correlation on the largest scales in the sky - which are, of course, the quadrupole and octupole.
\subsection{WMAP Map analysis}
	
We now examine WMAP's seven year maps in our analysis \citep{WMAP7}.  We use the full sky and cut-sky ILC (Internal Linear Combination)
maps\footnote{We have checked these results for robustness by comparing to a
  co-added foreground-cleaned Q-V-W band map, and the results there are very
  similar to that of the cut-sky ILC.}.  The maps are first degraded to
$\nside=64$ (corresponding to pixel scale $\simeq 1\degr$); next, the KQ75 mask
is applied, and finally all resulting pixels that are more than 10\% masked
(i.e.\ with mask value of less than 0.9)
are excluded from the analysis. Finally, the dipole was removed.

From inspection of the full-sky ILC $S_{1/2}(\hat{\Omega})$ 
(Fig.~\ref{fig:QORegions}, right panel), there is an obvious correlation between the magnitude of
$S_{1/2}(\hat{\Omega})$ contributions and the regions where the quadrupole and octupole have power, as expected from our analysis of random maps. Quantifying this result using our previous definition of the QO region, we find that $S_{1/2} ^{\rm QO}$ = 8837 $(\mu \rm{K})^4$, while $S_{1/2} ^{\rm non-QO}$ = -349 $(\mu \rm{K})^4$.  \footnote{Note that
  $S_{1/2}^{\rm QO}$ can be
  greater than the full $S_{1/2}$, since the contributions 
from each individual $S_{1/2}(\hat{\Omega}_{\rm QO})$
can be negative as well as positive, and their 
sum over the full sky is equal to the full $S_{1/2}$.} The values observed in the full-sky map are very typical, compared to our random map results.

Moving on to the cut-sky map, there is much less power on large angular scales, as expected;  only 0.06$\%$ of random maps have a similarly low value of $S_{1/2}$. And as the empirical relationship predicts, $S_{1/2}^{\rm QO}$ is similarly low, at the 0.046$\%$ level. In comparison, the non-QO contribution, $S_{1/2}^{\rm non-QO}$ is quite typical (15.6\% confidence level.)  Thus, the strong correlation that we observed in random maps continues to hold for the anomalous cut-sky map, as shown in Figure \ref{fig:SQO}. What is unlikely about our sky is simply for $S_{1/2}^{\rm QO}$ to exhibit such a low value in the first place.

It is known that the quadrupole in the CMB sky is lower than
expected from $\Lambda$CDM results, though not anomalously so.\citep{Komatsu:2010} It makes sense
to consider what happens to our results above if we use the $\Lambda$CDM theoretical power
spectrum for Monte Carlo simulations as before, but substituting the low WMAP
quadrupole ($C_2 = 210\; (\mu K)^2$), in order to test if our result is still valid. As expected, the
significance of $S_{1/2}$ relative to these new synthetic maps is now lower
($0.70\%$ compared to $0.06\%$ before), and a similar effect is seen in our
new statistics: 
$S_{1/2}^{\rm QO}$ has a significance of $0.58\%$ (previously $0.046\%$), and
$S_{1/2}^{\rm non-QO}$ is again typical -- $27.4\%$ C.L., compared to $15.6\%$
before.  Therefore, comparison to Gaussian random skies conditioned to have
low $C_2$ gives results that are somewhat less significant then before, which
is expected since we already knew that part of the low-$S_{1/2}$ problem
originated from low quadrupole. However, the strong correlation observed in random maps continues to hold. In addition, the various statistics are still low
with $\leq 1\%$ probability, even compared to skies with the low quadrupole,
and the lowness is once again due to lack of power from the quadrupole and
octupole regions.
\section{Conclusion}
In this paper, we first introduced the fixed-vertex correlation function $C(\theta, \hat{\Omega})$, which has the potential to be a useful tool for studying the statistical isotropy of the correlation function. We then observe that the value of $S_{1/2}$ in random skies is correlated with an expected anisotropy in the quadrupole and octupole regions of the sky. Our sky, known to have a low $S_{1/2}$, also receives low contributions from its quadrupole and octupole regions. We have a bit of a chicken and egg problem here, as we cannot tell if the low $S_{1/2}^{\rm QO}$ caused a low $S_{1/2}$, or vice versa, or if both were caused by an underlying factor. It is worth considering that this result and the one of \citet{PeirisPontzen} may stem from the same root cause.

It is also interesting that random Monte Carlo skies (which are generated from the assumption of statistical isotropy) and the reconstructed full-sky ILC (which assumes statistical isotropy in the reconstruction process) have similar values of $S_{1/2}$ and $S_{1/2}^{\rm QO}$, while the cut-sky temperature map is anomalous. These findings may provide a hint for a future understanding of the origin of large-angle missing power in the CMB.
\section{Acknowledgements}
We thank Dragan Huterer for initial guidance in the project and many useful suggestions. We thank Glenn Starkman for introducing the idea of $C(\hat{\Omega},\theta)$. We thank Hiranya Peiris for a useful communication. We acknowledge use of the HEALPix package \citep{healpix} and the Legacy Archive for Microwave Background Data Analysis (LAMBDA). This work was partly supported by NSF under contract AST-0807564, NASA under contract NNX09AC89G, and the University of Michigan.

\end{document}